\documentstyle[12pt]{article}
\def\lsim{\mathrel{\lower4pt\hbox{$\sim$}}
\hskip-13pt\raise1.6pt\hbox{$<$}\;}

\def\gsim{\mathrel{\lower4pt\hbox{$\sim$}}
\hskip-13pt\raise1.6pt\hbox{$>$}\;}

\begin{document}

\begin{flushright}
SLAC-PUB-6524\\
hep-ph/9406391 \\
June  1994\\
T
\end{flushright}
\begin{center}
\large\bf
 CP Violation in  $B^\pm\to \gamma\pi^\pm\pi^+\pi^-$\ \  *
\normalsize  \rm
\bigskip

D. Atwood$^{\rm a}$ and A. Soni$^{\rm b}$ \\
\bigskip
\end{center}

\begin{flushleft}

a) Stanford Linear Accelerator Center,
Stanford University, Stanford, CA\ \ 94309, USA \\
b) Department of Physics, Brookhaven National
Laboratory, Upton, NY\ \ 11973, USA\\

\end{flushleft}
\bigskip

\bigskip

\small
\begin{center}
{\bf Abstract}
\end{center}

We  consider CP violating effects in  decays of the type $B^\pm\to \gamma
a^\pm_{1,2}$ where $a_{1,2}$ are the $J^P=1^+$ and $2^+$ resonances each
decaying to the common final state via $a^\pm_{1,2} \to\pi^\pm\rho^0$.
The resonances enhance the CP asymmetries and also knowledge of their
masses and widths facilitates calculations of the effects. Several types
of CP asymmetries are sizable ($\sim10$--30\%) requiring about
(3--10)${}\times10^8$ $B^\pm$ mesons for detection at the $3\sigma$
level thereby providing a method for measuring the angle $\alpha$ in the
unitarity triangle. \bigskip

\vfill

\begin{center}

Submitted to {\it Physical Review Letters}

\end{center}

\vfill
\hrule
\vspace{5 pt}
\noindent
* Work supported by the Department of Energy Contracts
DE-AC03-76SF00515 (SLAC) and DE-AC02-76CH0016 (BNL).

\vfill
\eject

\normalsize
\bigskip\bigskip\bigskip

Unlike the neutral $B$ system,$^1$ wherein for several decay modes
predictions for CP asymmetries can be made with considerable confidence,
in charged $B$ decays reliable quantitative predictions for CP violation
are very difficult to make due to the traditional problems in
calculating hadronic matrix elements. To alleviate this outstanding
problem we consider final states that are dominated by at least two
neighboring resonances.$^2$ This has the advantage that, to the extent
that the resonances dominate the channels, the known widths and masses
of the resonances give a crucial handle on reliably calculating the CP
violating asymmetries. Furthermore, dominance of the channels by the
resonances and coherent superposition of the contributing amplitudes
from the resonances can lead to significant enhancements in the
asymmetries.$^2$ Let us also briefly recall, in passing, that the
charged $B$ meson system has the advantage that 1) all CP violation is
unambiguously of the ``direct'' type, 2) no tagging of ``the other'' $B$
is necessary and 3) experiments can be performed at the conventional
machines (e.g.\ CESR) as well as at the asymmetric $B$-factories that
are under construction at SLAC and at the KEK\null.

We are thus led to  investigate the prospects for CP violation in
radiative decays of $B^\pm$ mesons to pionic final states, i.e.
$B^\pm\to \gamma\pi^\pm\pi^+\pi^-$. The key feature of this reaction
that we wish to exploit is in the region where it is dominated  by two
overlapping resonances, namely, the $J^{P} = 1^+$, $a_1$ ($M_{a_1} =
1260$ MeV, $\Gamma_{a_1} \sim400$ MeV) and $J^{P}=2^+$,
 $a_2$ ($M_{a_2}=1318$ MeV, $\Gamma_{a_2}=110$ MeV). So the reactions of
interest are:

\begin{eqnarray}
B^\pm\to\gamma a^\pm_1 \quad &, \quad a^\pm_1\to \rho^0 \pi^\pm \quad &,
\quad \rho^0\to\pi^+\pi^- \\
B^\pm\to\gamma a^\pm_2 \quad &, \quad a^\pm_2\to \rho^0\pi^\pm \quad &,
\quad \rho^0\to\pi^+\pi^-
\end{eqnarray}

\noindent The formalism for assessing CP violation effects in presence
of  interfering resonances was given in Ref.~2 where, as an
illustration, it was used for radiative decays of $B$-mesons to final
states that are dominated by kaonic resonances, i.e. $B\to\gamma K^\ast
(892)$, $\gamma K_1 (1270)$, $\gamma K_1 (1400)$, $\gamma K^\ast (1410)$
and $\gamma K_2 (1430)$.
This class of reactions are, of course, driven largely by the $b \to s$
penguin transition whereas what we will report in the present study are
purely pionic final states which therefore result from $b \to d$ quark
transitions. Since in the Standard Model (SM) all CP violation has to
proceed via a single, unique, invariant quantity$^3$ and since $b \to d$
transitions are  relatively suppressed compared to  $b \to s$, it is
therefore clear that CP violating asymmetries should be larger in
reactions of the type (1--2) compared to our previous study involving $B\to
\gamma K^\ast$-like resonances.

These reactions receive contributions from the penguin and the annihilation
graph as well. However, since due to the Cabibbo angle the annihilation
graph for $b\to d$
reactions is larger than it is for the reactions $b\to s$, the two
contributing graphs (namely the penguin and the
annihilation) tend to become of comparable strength and that too
enhances the prospects for larger CP asymmetries for reactions (1--2).
Indeed, asymmetries are typically several tens of percents so
that effects at the $3 \sigma$ level should be observable with about
$5 \times 10^8$ $B^\pm$ mesons. Furthermore, such a final state is expected to
reveal  CP-conserving asymmetries as well which depend on the CP
conserving ``interaction'' phase(s) originating from strong interactions
thus   giving a better handle on deducing the underlying CP-violating
CKM phase.

Since resonances $a_1$ and $a_2$ have different quantum numbers the
amplitudes for reactions (1) and (2) can be simply written as:

\begin{equation}
M_j = A_j \Pi_{j} b_j
\end{equation}

\noindent with $j=1,2$. Here $A_j$ describes the weak decay $B \to
\gamma a_j$ and therefore contains the CP-violating CKM phase. $\Pi_{j}$
is the Breit-Wigner propagator:

\begin{equation}
\Pi^{-1}_{j} = s - m^2_{j} + i \Gamma_{j} m_j \end{equation}

\noindent and thus is one source for the CP-conserving  ``interaction
phase''. In eqn.~(3) $b_j$ describes the strong decay of the resonance
$a_j$ to the final state $\rho_0 \pi^\pm$. Due to its width the decay of
the $\rho_0$ via $\rho^0 \to \pi\pi$ introduces an additional source of
an interaction  phase that has to be included.

As in Ref.~2 we use a bound state model$^{4,2}$  to describe the
conversion from the quark level weak amplitudes to the formation of
resonances in the exclusive channels via $B \to \gamma a_{(1,2)}$. We
thus find that the formation of $a_2$ via the annihilation graph is
extremely small and we consequently approximate it to zero. In addition,
using$^{6,7}$ $BR (b \to s \gamma) = 2.5 \times 10^{-4}$ (corresponding
to $m_t\sim170$ GeV), and the constraints from experiment and theory on
$b\to u$ and $b\to c$ transitions, $K$-$\bar K$ and $B$-$\bar B$
mixing$^{8,9}$ we find:

\begin{eqnarray}
BR(B\to\gamma a_1)_{\rm pen} &\equiv B^{\rm pen}_1 & \simeq
(1.3\hbox{--}2.0) \times 10^{-7} \qquad (a) \nonumber\\
BR(B\to\gamma a_1)_{\rm ann} &\equiv B^{\rm ann}_1 & \simeq
(1.5\hbox{--}4.6) \times 10^{-7} \qquad (b) \\
BR(B\to\gamma a_2)_{\rm pen} &\equiv B^{\rm pen}_2 & \simeq
(1.0\hbox{--}1.7) \times 10^{-7} \qquad (c) \nonumber
\end{eqnarray}

\noindent The CP-violating phase $\delta_{cp}$ is then given by:

\begin{equation}
\delta_{cp} = {\rm Arg}\; \left[ A^{\rm pen}_2 (A^{\rm ann *}_1 + A^{\rm
pen*}_1 ) \right]
\end{equation}

\noindent Using the standard Wolfenstein parameterization$^{11,12,8}$ of
the CKM matrix one gets:

\begin{eqnarray}
{\rm Arg}\; (A^{\rm pen}_2 A^{\rm ann*}_1) &=& {\rm Arg}\;[ (\rho+i\eta)
(1-\rho+i\eta)] \\
&=& \gamma + \beta = \pi - \alpha
\end{eqnarray}

\noindent where $\rho$, $\eta$ are the usual parameters of that matrix
and $\alpha$, $\beta$ and $\gamma$ are the angles in the unitarity
triangle.$^{12,8}$ Thus

\begin{equation}
\delta_{cp} = {\rm Arg}\; \left[ \sqrt{B^{\rm pen}_1} -  \sqrt{B^{\rm
ann}_1} e^{-i\alpha} \right] \end{equation}

\noindent and therefore it follows that the charged $B$-mesons via modes
under discussion, namely (1,2) should allow a determination of one of
the angles (namely $\alpha$) in the unitarity triangle. Note also that
as these {\it branching ratios\/} get experimentally measured (which
should happen well before the CP asymmetries become observable), the
uncertainties in equation (9) due to the model dependence of equation
(5) should get significantly reduced.

For the strong decay $a_1\to3\pi$ the amplitude is given by

\begin{equation}
b_1=c_1m_1a^\mu_1 [ (p_0-p_1)_\mu\pi_{01}+ (p_0-p_2)_\mu \pi_{02}]
\end{equation}

\noindent where $m_1$ is the mass of $a_1$, $p_1$, $p_2$ are the momenta
of the two identical pions and $p_0$ that of the third pion,
$\pi_{ij}=[(p_i+p_j)^2-m^2_\rho + i\Gamma_\rho m_\rho]^{-1}$ and
$i,j=0,1,2$. Similarly, for $a_2\to3\pi$ the strong amplitude is

\begin{equation}
b_2= 2 c_2 a^{\mu\nu}_2 [(p_0-p_1)_\mu p_{2\nu} \pi_{01} +
(p_0-p_2)_\mu p_{1\nu}\pi_{02}]
\end{equation}

\noindent The constant $c_1$ and $c_2$ are determined by the measured
total widths$^{13}$ to be 22.75 and 28.20 respectively.

Contributions to CP-violating observables require interference between
the CP-violating phase $\delta_{\rm CP}$ with the strong rescattering
phase(s). In our formulation, encapsulated in equation~(3), the strong
phases originate from the widths of $a_{1,2}$ as well as from the width
of $\rho_0$. To the extent that these resonances dominate the final
states, the theoretical difficulties in calculating the interaction
phases are bypassed as the knowledge gained from the existing
experimental information$^{13}$ of the widths and masses of the
resonances suffices.

To understand the various asymmetries that arise we rewrite the
propagators for $a_{1,2}$ so that the relevant rescattering phases are
explicitly exhibited. Thus for the $a_{1,2}$ we write:

\begin{equation}
 \Pi_j =\hat\Pi_j \exp (-i\alpha_j)
\end{equation}

\noindent Furthermore, since there are two pions with the same charge in
the final state (e.g.\ $B^+\to\gamma\pi^+ (p_1)+ \pi^+(p_2) +
\pi^-(p_0)$), therefore there are two ways in which the $\rho$
propagator enters. For convenience, we decompose this in a symmetric
($\Sigma$) and an antisymmetric ($\Delta$) combination:

\[ \Sigma=\pi_{02} + \pi_{01} \quad ; \quad \Delta= \pi_{02} -\pi_{01}
\]

\noindent Once again we factor out the phases

\[ \Sigma=\hat\Sigma\exp (-i\rho_1) \quad ; \quad \Delta=\hat \Delta
\exp (-i\rho_2) \]

\noindent The resulting phases that determine the asymmetries are then
the differences:

\[ \Delta\alpha = \alpha_1-\alpha_2 \quad \mbox{and} \quad \Delta\rho =
\rho_1-\rho_2 \]

Altogether there are six types of CP violating asymmetries that arise.
All of the CP-odd quantities, of course, have to be proportional to
$\sin\delta_{\rm CP}$. But, in addition, those observables that are odd
under ``naive time-reversal'' (denoted by $T_N$ and meaning
time${}\to-{}$time
without interchange of initial and final states) will also have to be
proportional to $\cos\Delta\alpha$ or $\cos(\Delta\alpha\pm\Delta\rho)$
whereas the $T_N$-even ones are proportional to $\sin\Delta\alpha$ or
$\sin(\Delta\alpha\pm\Delta\rho)$. Thus the square of the invariant
amplitude can be expressed as:

\begin{equation}
|M_1+M_2|^2 = P+\sin\delta_{\rm CP} R
\end{equation}

\noindent where $P$ is the CP conserving part and $R=(R_o + R_e)$ is the
CP violating part. Here $R_o$ (i.e.\ the $C$-even, $P$-odd, $T_N$-odd
part) contains terms proportional to $\cos\Delta\alpha$ or
$\cos(\Delta\alpha \pm \Delta\rho)$. $R_e$ (i.e.\ $C$-odd, $P$-even,
$T_N$-even part) contains terms proportional to $\sin\Delta\alpha$ or
$\sin(\Delta\alpha\pm \Delta\rho)$.

Numerical results for the asymmetries are given in Table~1.$^{14}$ A
simple observable that exhibits a sizable asymmetry is

\begin{equation}
\epsilon_{fb} = \langle Q_B \sigma(\cos\theta)\sigma(s-s_0)\rangle
\end{equation}

\noindent where $\sigma(x)=+1$ if $x>0$ and $-1$ if $x<0$, $\cos\theta
\equiv \hat p_0\cdot \hat q$ where,  $\vec q$ is the momentum of the
photon and $\vec p_0$ is the momentum of the $\pi^-$ ( in $B^+$ decay)
in the rest frame of $a_{1,2}$. $Q_B$ is the charge of the $B^\pm$
meson.
The quantity $s$ is the invariant mass of the three pions
and
\begin{equation}
s_0={\Gamma_1m_1m_2^2-\Gamma_2m_2m_1^2
\over
\Gamma_1m_1-\Gamma_2m_2}
\end{equation}
is the point at which $\sin\Delta\alpha$ switches sign.
Thus $\epsilon_{fb}$ is a CP-violating forward-backward asymmetry
and from
Table~1 we see that it ranges from 7-11\%.

In the Table we also show a simple triple product correlation asymmetry

\begin{equation}
\epsilon_t \equiv \langle \sigma (\sin2\phi)\rangle
\end{equation}

\noindent where $\sin\phi = [(\vec p_2\times\vec p_1) \cdot \vec
q]/|\vec p_1\times\vec p_2|\; |\vec q|$; $\cos\phi=(\vec p_2-\vec p_1)
\cdot \vec q/ |\vec p_2-\vec p_1|\; |\vec q|$. For the purpose of this
observable the momentum of the identical pions ($p_{1,2}$) are ordered
by energy. The resulting CP violating asymmetry ranges from 7 to 10\%.

{}From eqn.~(11), following Ref.~15, the optimal observable for
CP-violation is

\begin{equation}
\epsilon_{opt} \equiv \langle R/P\rangle \end{equation}

\noindent We find $ \epsilon_{opt}$ to be about 20--35\%. This
CP violating observable can be separated into $T_N$-odd and $T_N$-even
pieces. The corresponding observables, $\epsilon_o \equiv \langle
R_o/P\rangle$ and $ \epsilon_e = \langle R_e/P\rangle$ are about
15--20\% and 20--30\% respectively.

In addition to such CP violating asymmetries, the final state also
exhibits rather interesting CP conserving asymmetries. As an example of
this class of asymmetries we show in Table~1:

$$\zeta_{fb} \equiv \langle\sigma(\cos\theta)\rangle $$

\noindent which is about 20--25\%.
Measurements of such CP conserving
asymmetries would be helpful in pinning down the CP-conserving
interaction phase(s).

In Figure 1 we show the differential asymmetries as a function of $s$
for the three cases  mentioned above.  We have assumed  typical values
for the CKM parameters.

In calculating the numbers given in Table~1 and in Fig.~1 we used the
bound state model of Isgur {\it et al}$^4$ with modifications given in
Ref.~2. The ranges in Table~1 are obtained by varying over the allowed
90\%CL limits of the CKM parameters.$^9$ We note, in passing, that the
asymmetries, being ratios of rates, tend to be less dependent on the
bound state model as compared to the rates. Also, as we mentioned
earlier, the model dependence should be further reduced as data on
branching fractions becomes available.

As the numbers in the Table indicate these effects should be observable
with about $10^8$--$10^9$ $B^\pm$ mesons. This is especially notable
given that we are dealing here with radiative transitions. The basic
idea of interfering resonances when used in the context of purely
hadronic modes should need significantly fewer $B$ mesons. We shall
discuss some of these applications in forthcoming publications.

\bigskip\bigskip
\leftline{\bf References}
\bigskip

\begin{enumerate}

\item For recent reviews see: H. Quinn, SLAC-PUB-6438; R. Peccei,
preprint \#Hep-ph-9312352.

\item D. Atwood and A. Soni, SLAC-PUB-6425, Jan.~1994.

\item C. Jarlskog, Phys. Rev. Lett {\bf 55}, 1039, 1985; C. Jarlskog
and R. Stora, Phys.\ Lett.\ B{\bf208}, 268 (1988);
see also, L. L. Chau and W. Y. Keung, Phys. Rev. Lett {\bf 53}, 1802
(1984).

\item N. Deshpande, P. Lo and J. Trampetic, Z. Phys.\ C{\bf40}, 369
(1988); N. Deshpande, P. Lo, J. Trampetic, G. Eilam and P. Singer,
Phys.\ Rev.\ Lett.\ {\bf59}, 183 (1987). See also N. Isgur, D. Scora, B.
Grinstein and M. Wise, Phys.\ Rev.\ D{\bf39}, 799 (1989). We note that
predictions for $B\to K^\ast\gamma$ from these models are in rough
agreement with the lattice calculations of Ref.~5. Note also that we
used another bound state model described in Ref.~2 and found the
difference in the predictions between the two models of the rates given
in Eqn.~(5) to be less than 50\%.

\item {} C. Bernard {\it et al}. Phys.\ Rev.\ Lett.\ {\bf72}, 1402
(1994); see also K.C. Bowler {\it et al}., Phys.\ Rev.\ Lett.\ {\bf72},
1398 (1994).

\item  B.~Grinstein, R.~Springer, and M.~Wise, Nucl.\ Phys.\
{\bf B339}, 269 (1990);
R.~Grigjanis, P.~O'Donnell, M.~Sutherland, and
H.~Navelet, Phys.\ Lett.\ {\bf B237}, 252 (1990);
M.~Misiak, Phys.\ Lett.\ {\bf B269}, 161(1991);
Nucl.\ Phys.\ {\bf B393}, 23 (1993); A.J. Buras, M. Misiak, M. Munz and
S. Pokorski, preprint MPI-PH-93-77.

\item For the related experimental literature see: R. Ammar {\it et
al}., CLEO Collaboration, Phys.\ Rev.\ Lett.\ {\bf71}, 674 (1993); Y.
Rozen, PhD thesis, Syracuse University (1993).

\item For a recent update on the CKM parameters and the unitarity
triangle see M. Witherell, UCSB-HEP-94-02.

\item Our constraints are a little different from those in Ref.~8. The
difference is primarily due to the fact that we use results from
the lattice for the relevant hadronic parameters. Thus we use (in
standard notation), $B_K=.8\pm.1$, $f_B=187\pm50$ MeV and $B_B=1\pm.2$,
where the errors are our best estimates at 90\% CL\null.
As a result we
find
$.15\lsim|V_{td}/V_{ts}|\lsim.30$,
$10^\circ\lsim|{\rm Arg} V_{td}|\lsim 40^\circ$;
$.035\lsim |V_{ub}/V_{cb}|\lsim .130$
and
$30^\circ\lsim|{\rm Arg} V_{ub}|\lsim 135^\circ$.
See also Ref.~10.

\item More details of this work will be published in a forthcoming
article.

\item See Particle Data Group, Phys.\ Rev.\ D{\bf11} (1992),
p.III.65-67; L.~Wolfenstein, Phys.\ Rev.\ Lett.\ {\bf 51}, 1945 (1983);
see also, L. L. Chau and W. Y. Keung, {\it ibid}.

\item See e.g. Y. Nir and H. Quinn, Ann.\ Rev.\ Nucl.\ Part.\ Sci.\
{\bf42}, 211 (1992).

\item See Particle Data Group, {\it ibid}, in particular
p. II.7.

\item Since the interfering resonances do not have identical quantum
numbers partial rate asymmetries cannot arise. See Ref.~2.

\item D. Atwood and A. Soni, Phys.\ Rev.\ D{\bf45}, 2405 (1992).

\end{enumerate}

\bigskip
\bigskip
\newpage

\noindent {\bf Table 1}: Observables and their transformation
properties. The ranges of the expected asymmetries are obtained by
varying over the allowed region of the CKM parameters. (see Ref.~9).
$N^{3\sigma}_B$ is the number of $B^\pm$ needed for detection at the
$3\sigma$ level. \bigskip \begin{center}  \begin{tabular}{|l|c|c|c|c|c|}
\hline Observable & \multicolumn{3}{c}{Transformation Property} &
Expected & $N^{3\sigma}_B/10^8$ \\
\cline{2-4} & CP & $P$ & $T_N$ & Size & \\
\hline
$\epsilon_{fb}$ & $-$ & $+$ & $+$          & 7--11\%  & 30--40 \\
$\epsilon_t$ & $-$ & $-$ & $-$             & 7--10\%  & 40--50 \\
$\epsilon_{\rm opt}$ & $-$ & Mixed & Mixed & 20--35\% & 3--5   \\
$\epsilon_e$ & $-$ & $+$ & $+$             & 20--30\% & 5--6   \\
$\epsilon_o$ & $-$ & $-$ & $-$             & 15--20\% & 8--12  \\
$\zeta_{fb}$ & $+$ & $+$ & $+$             & 20--25\% & 4--10  \\ \hline
\end{tabular}
\end{center}

\bigskip\bigskip\bigskip\bigskip
\leftline{\bf Figure Captions:}
\bigskip

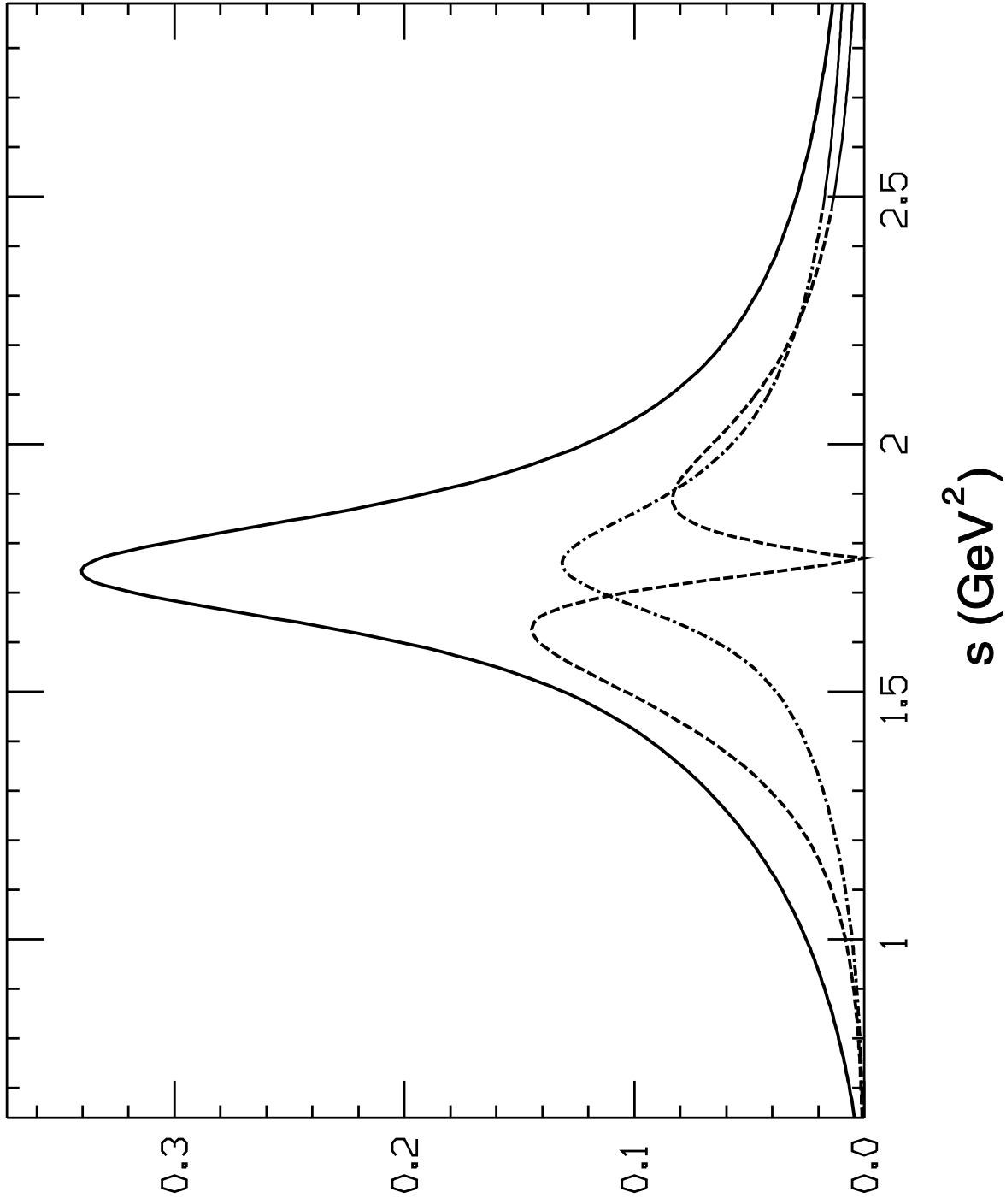
\leftline{ Figure 1:}
Asymmetries as a function of $s$
for the Wolfenstein parameters
$\{A=.86,\rho=.10,\eta=.45\}$.
The solid line is for
$|m_1^2{d \zeta_{fb}\over d s}|$;
the dashed line for
$|m_1^2{d \epsilon_{fb}\over d s}|$
and
the dot-dashed line is for
$|m_1^2{d \epsilon_{t}\over d s}|$.

\end{document}